\documentclass[
    aps,
    prb,
    reprint,
    superscriptaddress,
    longbibliography,
    nofootinbib
]{revtex4-2}

\usepackage{amsmath,amssymb,bm}
\usepackage[colorlinks=true,linkcolor=blue,citecolor=blue,urlcolor=blue]{hyperref}
\usepackage{xcolor}
\newcommand{\sus}{\smash[b]{\chi}}

\begin{document}

\title{\texorpdfstring{Unified contact-free formulation of linear-response transport theory}{Unified contact-free formulation of linear-response transport}}
\author{Ziqian Wang}
\affiliation{International Center for Quantum Materials, School of Physics, Peking University, Beijing, China}
\author{Ji Feng}
\email{jfeng11@pku.edu.cn}
\affiliation{International Center for Quantum Materials, School of Physics, Peking University, Beijing, China}
\affiliation{Hefei National Laboratory, Hefei, China}

\date{\today}

\begin{abstract}
Thermal Hall transport is conventionally formulated as a current--current Kubo response supplemented by an energy-magnetization correction obtained through an auxiliary pseudogravitational magnetic field. Motivated by this structure, we develop a unified contact-free moment formulation of linear response to scalar sources. For channels whose source contributions vanish in the dc limit, the transport coefficient is the low-frequency residue of a retarded moment--moment correlator, with equal-time endpoint contributions incorporated before the dc limit is taken. The formulation uses only scalar sources, requires no separate magnetization subtraction, is independent of the local-current gauge, and is invariant under admissible redistributions of bond or interaction energy. It organizes particle, grand-energy, and specified spin-density channels within a common source--moment response matrix. Quadratic Landau--Lifshitz spin waves illustrate the method: the secular growth of polarization-moment correlations isolates the finite Hall residue and yields the magnon thermal Hall and spin Nernst coefficients from the same paraunitary band geometry. The formulation provides a unified operator framework that simplifies the organization of particle, thermal, and spin transport and can streamline calculations by eliminating separate contact and magnetization corrections.
\end{abstract}

\maketitle

\section{Introduction}

Thermal Hall transport has become an important diagnostic of quantum geometry in both electronic and charge-neutral systems, ranging from quantum Hall fluids and topological bands to magnetic insulators, superconductors, and collective bosonic excitations.\cite{KaneFisher1997,Banerjee2017Nature,Banerjee2018Nature,ReadGreen2000,KatsuraNagaosaLee2010,Onose2010,Hirschberger2015Science,Strohm2005,ZhangRenWangLi2010,MatsumotoMurakami2011,MatsumotoMurakami2014,Kasahara2018Nature} Luttinger's fictitious gravitational potential provides the standard scalar source for thermal transport by coupling to the energy density.\cite{Luttinger1964} It converts a temperature gradient into a mechanical perturbation and thereby makes linear-response theory applicable to heat currents.\cite{Kubo1957,KuboBook}

Thermal gradients can also generate transverse spin responses in magnetic insulators. In the magnon spin Nernst effect, Berry curvature deflects opposite-spin magnon branches in opposite transverse directions, producing a spin-current or edge-spin response even when the net thermal Hall signal cancels.\cite{Cheng2016,Zyuzin2016,Kovalev2016} These works provide the spin-transport counterpart of magnon thermal Hall physics, formulated in terms of magnon Berry curvature and spin-current signals. Together, the magnon thermal Hall and spin Nernst effects point to a broader class of transverse responses driven by scalar thermodynamic forces.

A special subtlety of thermal transport is the role of energy magnetization. Cooper et al. emphasized that local-equilibrium currents include magnetization currents, which circulate within the sample and should not be counted as transport currents through its boundary.\cite{Cooper1997} In that formulation, the energy magnetization is exposed by introducing an auxiliary pseudogravitational magnetic field. Qin et al. subsequently used this construction to obtain the magnetization correction required for thermal Hall transport.\cite{Qin2011} This insight underlies much of the modern theory of thermal Hall response in electronic and bosonic systems. The conventional route is physically correct, but it introduces a second fictitious gravitational probe and separates the final response into a Kubo part and a magnetization part. The two pieces depend on the chosen local energy and current representatives, whereas only their sum is a transport observable.

The present work develops a moment formulation of scalar-source transport, with thermal transport as the motivating case. The source conjugate to a uniform thermal force is the grand-energy polarization moment. Its time derivative gives the integrated heat current, and the dc coefficient follows from the low-frequency response of the corresponding moments. Particle and specified spin-density sources enter the same source--moment basis. In this organization, the equal-time endpoint that would otherwise appear as a source-contact or magnetization correction is absorbed before the transport limit is taken. The resulting formula is contact free: it contains no explicit additive magnetization term while still yielding the magnetization-corrected transport response.

The construction has several useful consequences. First, because it is formulated in terms of density moments, it does not require an independent choice of local-current gauge. Second, the moment formula is invariant under improvements of local conserved densities, including the lattice freedom to assign bond or interaction energies to different sites. Third, heat magnetization is not introduced as an independent response: the formulation uses only Luttinger's scalar gravitational potential and the grand-energy polarization generated by that scalar source. The endpoint/contact cancellation then reproduces the magnetization-corrected transport coefficient without a pseudogravitational magnetic field. Once the relevant spin-density components and their sources are specified, the same source--moment logic organizes a unified particle, thermal and spin response matrix.

The rest of the paper is organized as follows. We first introduce density moments and relate them to integrated currents and sources, while distinguishing the density, source, and local-current ambiguities. We then derive the contact-free moment Kubo formula, in which the equal-time contact correction is absorbed through an endpoint cancellation before the dc limit is taken. Its low-frequency structure is analyzed, separating the dc residue from the ballistic sector. Next, we connect the result to the conventional heat-magnetization description and establish invariance under density improvements. Finally, we apply the formula to quadratic Landau--Lifshitz spin waves. This example illustrates how the finite dc residue is isolated from the ballistic sector and how different transported densities are handled within the same paraunitary framework; the recovery of the magnon thermal Hall and spin Nernst formulas then provides a validation of the formulation.

\section{Contact-free transport coefficients}

This section develops the contact-free moment formulation. We first define finite-$\boldsymbol q$ moments of densities and sources and relate their first moments to integrated currents, thereby separating the density-representative, current--source, and local-current-gauge ambiguities. We then derive the scalar-source response and show that the explicit contact correction cancels the equal-time endpoint of the Kubo susceptibility, leaving the contact-free moment--moment kernel of Eq.~\eqref{eq:transport}. Its low-frequency structure separates the finite dc residue from the ballistic sector and establishes its relation to the Einstein--Helfand construction. Finally, we identify the grand-energy moment commutator with heat magnetization in a chosen current gauge and prove that the transport coefficient is invariant under admissible density improvements.

\subsection{Moment and current}

We begin by introducing density moments that are compatible with standard periodic boundary conditions. As we will show, the first moments are naturally related to the corresponding integrated currents and sources and will be used in the response calculations below.

Let $\hat a(\boldsymbol r)$ be the density of a quantity $a$. Its continuity equation in the presence of a source is
\begin{equation}
\dot{\hat a}(\boldsymbol r)
+
\partial_\alpha\hat j^a_\alpha(\boldsymbol r)
=
\hat\sigma^a(\boldsymbol r),
\label{eq:continuity}
\end{equation}
where $\hat j^a_\alpha(\boldsymbol r)$ is the local current density associated with $a$ in the $\alpha$ direction, and $\hat\sigma^a(\boldsymbol r)$ is the corresponding local source density. In the applications considered here, $a$ may represent particle number $N$, energy $E$, grand energy $K$, or a specified spin component $S^\nu$. For a conserved quantity, $\hat\sigma^a=0$, whereas for a nonconserved spin component, $\hat\sigma^{S^\nu}$ is the corresponding spin-torque density. The corresponding integrated current is
\begin{equation}
\hat J^a_\alpha
=
\int \mathrm d\boldsymbol r\,
\hat j^a_\alpha(\boldsymbol r).
\end{equation}

We assume that the external fields are weak and slowly varying in space and may therefore be treated within linear response. The grand Hamiltonian in the presence of the external potentials $\phi^a$ is written as
\begin{equation}
\hat K[\phi]
=
\int \mathrm d\boldsymbol r\,
\big[
\hat h(\boldsymbol r)
-
\mu_0\hat n(\boldsymbol r)
+
\sum_a
\phi^a(\boldsymbol r)\hat a(\boldsymbol r)
\big],
\end{equation}
where $\hat h$ is the bare Hamiltonian density, $\mu_0$ is the uniform reference chemical potential, and $\phi^a$ is the external field conjugate to the density $\hat a$. We denote the grand-energy density by
\begin{equation}
\hat k(\boldsymbol r)
=
\hat h(\boldsymbol r)-\mu_0\hat n(\boldsymbol r).
\end{equation}
For example, when $\hat a=\hat n$, $\phi^a$ represents a perturbation of the electrochemical potential; when the thermal channel is included, one takes a source channel $a=K$ whose density is the grand-energy density $\hat k$, with $\phi^K$ the Luttinger gravitational potential and $\hat{\boldsymbol j}^{\,K}$ the heat-current density; and when $\hat a=\hat s^\nu$, $\phi^a$ is a Zeeman field along the $\nu$ direction. The sign relating $\phi^a$ to the corresponding physical field is absorbed into the definition of $\phi^a$.

For a spatially distributed operator $\hat f(\boldsymbol r)$, we define its moments using a finite-$\boldsymbol q$ regulator:
\begin{equation}
\hat F_{\alpha\cdots\beta}
=
\left.
(i\partial_{q_\alpha})\cdots
(i\partial_{q_\beta})
\hat f(\boldsymbol q)
\right|_{\boldsymbol q=0},
\label{eq:moment}
\end{equation}
where we adopt the Fourier convention
$
\hat f(\boldsymbol q)
=
\int \mathrm d\boldsymbol r\,
e^{-i\boldsymbol q\cdot\boldsymbol r}
\hat f(\boldsymbol r).
$
Applying this definition to $\hat a(\boldsymbol r)$ and $\hat\sigma^a(\boldsymbol r)$ gives the corresponding moments $\hat A_{\alpha\cdots\beta}$ and $\hat\Sigma^a_{\alpha\cdots\beta}$. The zeroth-moment equation is
\begin{equation}
\dot{\hat A}
=
\hat\Sigma^a,
\end{equation}
where $\hat A$ is the total amount of $a$, and $\hat\Sigma^a$ is its total production rate. At first order, Eq.~\eqref{eq:continuity} gives the \emph{bare} integrated current
\begin{equation}
\hat J^a_\alpha
=
\dot{\hat A}_\alpha
-
\hat\Sigma^a_\alpha.
\label{eq:int-current}
\end{equation}
Thus, the integrated current is the time derivative of the polarization moment with the first source moment subtracted. In the source-free case, Eq.~\eqref{eq:int-current} reduces to
\begin{equation}
\hat J^a_\alpha
=
\dot{\hat A}_\alpha.
\end{equation}
Hardy gave a similar expression for the integrated heat current in terms of the first moment of the energy density.\cite{Hardy1963} The present formulation generalizes that result to arbitrary conserved or nonconserved densities and their associated sources.
The finite-$\boldsymbol q$ definition in Eq.~\eqref{eq:moment} allows these moment relations to be formulated without introducing an ill-defined position operator under periodic boundary conditions.

We take this opportunity to highlight three distinct ambiguities in the formulation of a local transport current that should not be conflated. The first concerns the choice of local density. A given extensive operator fixes only the spatial integral of its density, while the local density itself may be shifted by a total divergence without changing the total quantity under periodic boundary conditions. For energy transport, this freedom is commonly referred to as the energy-density ambiguity~\cite{Marcolongo2016,Ercole2016}. Once a particular microscopic Hamiltonian density, and hence the grand-energy density $\hat k=\hat h-\mu_0\hat n$, has been specified, this freedom is fixed at the level of the formulation. The associated moment operators, source operators, and response functions must then be constructed consistently from that choice.

A second ambiguity arises when the density is not conserved. The continuity equation determines only the combination $\partial_\alpha\hat j^a_\alpha-\hat\sigma^a$ and therefore does not, by itself, specify how local nonconservation is divided between current and source. In particular, the simultaneous transformations
\begin{equation}
\hat j^a_\alpha
\rightarrow
\hat j^a_\alpha+\hat p^a_\alpha,
\qquad
\hat\sigma^a
\rightarrow
\hat\sigma^a+\partial_\alpha\hat p^a_\alpha
\end{equation}
leave Eq.~\eqref{eq:continuity} unchanged. This issue is especially prominent in spin transport with spin--orbit coupling, where a specified spin component is not generally conserved and a spin-torque density appears in its continuity equation~\cite{Shi2006}. In the present formulation, both the density and source operators are specified microscopically. Their first moments then determine the integrated current through Eq.~\eqref{eq:int-current}, without requiring an independent choice of the longitudinal part of the local current.

Even after the density and source have been fixed, a third freedom remains. The local current may be shifted by a divergence-free contribution,
\begin{equation}
\hat j^a_\alpha
\rightarrow
\hat j^a_\alpha
+
\partial_\beta\hat m^a_{\alpha\beta},
\qquad
\hat m^a_{\alpha\beta}
=
-\hat m^a_{\beta\alpha}.
\label{eq:current-gauge}
\end{equation}
Such a transverse redefinition leaves both the continuity equation and its source unchanged. Under periodic boundary conditions, it also has vanishing spatial integral and therefore does not modify the integrated current. We refer to this freedom as the local-current gauge freedom.

Magnetization currents belong to this transverse sector. They describe local circulating flow rather than net transport through a periodic sample. Historically, this distinction became essential in thermoelectric and thermal Hall response because the current obtained directly from an equilibrium or Kubo calculation can contain circulating contributions that do not represent transport through the sample. Cooper et al. gave a systematic separation of transport and magnetization currents in thermoelectric response~\cite{Cooper1997}. Qin et al. subsequently formulated the corresponding energy-magnetization correction for thermal Hall transport~\cite{Qin2011}. These developments build on the general linear-response framework introduced by Kubo~\cite{Kubo1957} and Luttinger's gravitational-potential formulation of thermal transport~\cite{Luttinger1964}. We shall return below to the relation between this transverse-current freedom and the conventional heat-magnetization correction.

\subsection{Contact-free response}

Having fixed the density and source operators and eliminated the local-current ambiguity at the integrated level, we now construct the response to the conjugate external fields. To first order in the spatial gradients of the external fields, the perturbation to the grand Hamiltonian is
\begin{equation}
\delta\hat K
=
\sum_a \phi^a_\beta\hat A_\beta.
\label{eq:moment-perturbation}
\end{equation}
Here \(\phi^a_\beta\) is the uniform long-wavelength gradient of the scalar source \(\phi^a\) conjugate to the density \(\hat a\); equivalently, it is the generalized force conjugate to \(\hat A_\beta\), the first moment defined in Eq.~\eqref{eq:moment}. For the thermal channel, we choose \(\phi_\beta^K=-\partial_\beta T/T\).
Here and below, the lowercase channel labels $a$ and $b$ determine the corresponding density moments $\hat A_\alpha$ and $\hat B_\beta$, respectively.

To linear order in the generalized forces, the \emph{total} integrated current operator becomes
\begin{equation}
\hat J^a_\alpha[\phi]
=
\hat J^a_\alpha
+
\sum_b
\hat M^{ab}_{\alpha\beta}\phi^b_\beta
-
\delta_\phi\hat\Sigma^a_\alpha
+
O(\phi^2),
\label{eq:source-dependent-current}
\end{equation}
where
\begin{equation}
\hat M^{ab}_{\alpha\beta}
=
-\frac{i}{\hbar}
\big[
\hat A_\alpha,
\hat B_\beta
\big]
\label{eq:contact-operators}
\end{equation}
is the equal-time moment commutator, and
$\delta_\phi\hat\Sigma^a_\alpha$ denotes the part of the source moment that is linear in the applied fields. If the microscopic source operator has no explicit field dependence, this term is absent. More generally, it must be retained as part of the source specification.

The two field-induced modifications in Eq.~\eqref{eq:source-dependent-current}, namely the moment-commutator term and the source variation, will be referred to collectively as the contact correction. In linear-response theory, a contact term is an instantaneous contribution arising from the explicit dependence of the observable on the applied field, rather than from the retarded response of the state. The modification of the current operator therefore gives an instantaneous contribution analogous to the diamagnetic term in the Kubo formula for electrical conductivity~\cite{Mahan2000,Giuliani2005}. Such explicit field-dependent and equal-time contributions are commonly
referred to as contact terms~\cite{Bradlyn2012}.

The linear-response coefficient $L$ is defined by
\begin{equation}
\frac{1}{V}
\delta\langle\hat J^a_\alpha\rangle
=
\sum_b
L^{ab}_{\alpha\beta}\phi^b_\beta.
\label{eq:transport-coeff}
\end{equation}
The response of the unperturbed current operator follows from the
first-moment continuity equation. With $\omega^+=\omega+i0^+$, the
relevant retarded susceptibility is
\begin{equation}
\sus_{\dot A_\alpha,B_\beta}(\omega)
=
-\frac{i}{\hbar}
\int_0^\infty \mathrm dt\,
e^{i\omega^+t}
\big\langle
\big[
\dot{\hat A}_\alpha(t),
\hat B_\beta
\big]
\big\rangle_0.
\end{equation}
All susceptibilities below are understood to be retarded. Integration by
parts gives
\begin{equation}
\sus_{\dot A_\alpha,B_\beta}(\omega)
=
\frac{i}{\hbar}
\big\langle
\big[
\hat A_\alpha,
\hat B_\beta
\big]
\big\rangle_0
-
i\omega^+
\sus_{A_\alpha,B_\beta}(\omega).
\label{eq:integration-by-part}
\end{equation}
The endpoint term is canceled exactly by the expectation value of
Eq.~\eqref{eq:contact-operators}. The dc transport coefficient is therefore
\begin{equation}
\begin{aligned}
L^{ab}_{\alpha\beta}
&=
-\frac{1}{V}
\lim_{\omega\to0}
\big[
i\omega^+
\sus_{A_\alpha,B_\beta}(\omega)
+
\sus_{\Sigma^a_\alpha,B_\beta}(\omega)
\big]
\\
&\quad
-
\frac{\delta\langle\delta_\phi\hat\Sigma^a_\alpha\rangle_0}
{V\delta\phi^b_\beta}.
\end{aligned}
\label{eq:general-retarded-response}
\end{equation}
Equation~\eqref{eq:general-retarded-response} separates the universal endpoint cancellation from the genuinely source-dependent contributions. For a conserved channel, or more generally whenever the source terms vanish in the dc limit, it reduces to
\begin{equation}
L^{ab}_{\alpha\beta}
=
-\frac{1}{V}
\lim_{\omega\to0}
i\omega^+
\sus_{A_\alpha,B_\beta}(\omega).
\label{eq:transport-retarded-moment}
\end{equation}

To express this result in canonical form, we introduce the canonical correlation
$\langle\hat X;\hat Y\rangle_0
= \beta_0^{-1}
\int_0^{\beta_0}\mathrm d\lambda\,
\langle e^{\lambda\hat K_0}\hat X e^{-\lambda\hat K_0}\hat Y\rangle_0$,
with connected correlations understood below. Using the Kubo identity
\begin{equation}
\big\langle
\big[
\hat A_\alpha(t),
\hat B_\beta
\big]
\big\rangle_0
=
i\hbar\beta_0
\big\langle
\dot{\hat A}_\alpha(t);
\hat B_\beta
\big\rangle_0,
\label{eq:kubo-identity-used}
\end{equation}
and integrating by parts a second time, we obtain a static term and a
moment--moment contribution. The static term is regular under the
finite-$\boldsymbol q$ regulator and is annihilated by the explicit factor
of $\omega$ in the dc limit. This reduction assumes that the
static canonical susceptibility and the source-moment susceptibilities
remaining in Eq.~\eqref{eq:general-retarded-response} have no singular
zero-frequency projection after the finite-$\boldsymbol q$ moment has
been regulated. If a gapless, metallic, or symmetry-breaking sector
produces additional secular growth, the regulator and the explicit
source terms must be kept until that infrared sector has been separated.

Consequently, whenever the source
contributions in Eq.~\eqref{eq:general-retarded-response} vanish in the
dc limit,
\begin{equation}
L^{ab}_{\alpha\beta}
=
-\frac{\beta_0}{V}
\lim_{\omega\to0}
(\omega^+)^2
\int_0^\infty \!\!\mathrm dt\,
e^{i\omega^+t}
\big\langle
\hat A_\alpha(t);
\hat B_\beta
\big\rangle_0.
\label{eq:transport}
  \end{equation}
Equation~\eqref{eq:transport} is the central result of this work. Here \(\omega^+=\omega+i0^+\), with \(0^+\) implementing the usual adiabatic switch-on and retarded boundary condition. In practical numerical evaluations, \(0^+\) may be replaced by a finite \(\eta>0\), with the limit \(\eta\to0^+\) taken after the thermodynamic limit. The resulting retarded prescription fixes the singular denominators in the low-frequency expansion below. The overall sign follows from the convention in Eq.~\eqref{eq:moment-perturbation}. The operators $\hat A_\alpha$ and $\hat B_\beta$ are polarization moments rather than ordinary local observables in the thermodynamic limit, so the dc response is encoded in the long-time growth of their correlator rather than in a finite static susceptibility. We now make this extraction precise.

\subsection{Low-frequency structure}

Two limits enter Eq.~\eqref{eq:transport}, and their order matters. The
thermodynamic limit $V\to\infty$ is taken first. The derivative at
$\boldsymbol q=0$ in Eq.~\eqref{eq:moment} is understood in this limit,
when the allowed wavevectors become continuous; at finite volume, the
correlator is a discrete pole sum and does not define a smooth dc
transport coefficient. The dc limit $\omega\to0$ is taken afterwards,
with $\omega^+=\omega+i0^+$ providing the retarded prescription.

The low-frequency structure can be read from the long-time behavior of the canonical correlator. In a clean translationally invariant system, the component of a polarization moment associated with ballistic or conserved motion can grow linearly in time.
When the long-time behavior admits the expansion
\begin{equation}
\big\langle
\hat A_\alpha(t);
\hat B_\beta
\big\rangle_0
=
C_{2,\alpha\beta}^{ab}t^2
+
C_{1,\alpha\beta}^{ab}t
+
C_{0,\alpha\beta}^{ab}
+
r_{\alpha\beta}^{ab}(t),
\label{eq:asymptotic}
\end{equation}
where $r_{\alpha\beta}^{ab}(t)$ is bounded or decays sufficiently rapidly,
the identity
$\int_0^\infty \mathrm dt\,e^{i\omega^+t}t^n
=i^{n+1}n!/(\omega^+)^{n+1}$
gives
\begin{equation}
L^{ab}_{\alpha\beta}(\omega)
=
\frac{\beta_0}{V}
\big[
C_{1,\alpha\beta}^{ab}
+
\tfrac{2i}{\omega^+}C_{2,\alpha\beta}^{ab}
+
O(\omega)
\big],
\label{eq:low-frequency-moment}
\end{equation}
where $L^{ab}_{\alpha\beta}(\omega)$ denotes the finite-frequency
expression obtained by retaining ${(\omega^+)^2}$ in
Eq.~\eqref{eq:transport}. The coefficient
$C_{2,\alpha\beta}^{ab}$ determines the ballistic contribution. For a
diagonal response with Hermitian moments,
$\operatorname{Re}L^{aa}_{\alpha\beta}(\omega)$ contains the Drude term
$2\pi\beta_0 C_{2,\alpha\beta}^{aa}\delta(\omega)/V$.

Stationarity and symmetry of the canonical product imply
$C_{2,\alpha\beta}^{ab}=C_{2,\beta\alpha}^{ba}$ and
$C_{1,\alpha\beta}^{ab}=-C_{1,\beta\alpha}^{ba}$. In particular, for
$a=b$, the $t^2$ coefficient is symmetric under
$\alpha\leftrightarrow\beta$ and therefore drops out of the spatially
antisymmetric Hall response, whose finite dc value is determined by the
antisymmetric part of $C_{1,\alpha\beta}^{aa}$. For cross channels, the
same relations connect the coefficients obtained by interchanging the two
moment insertions. The analytic expansion \eqref{eq:asymptotic} is the clean-system structure: the diagonal dissipative response resides entirely in the $t^2$ (Drude) coefficient. With a finite relaxation rate $1/\tau$ the $t^2$ growth is cut off for $t\gtrsim\tau$ and broadens into a Drude peak of width $1/\tau$, while the regular diagonal coefficient appears as a $|t|$-type growth of the correlator; Eq.~\eqref{eq:transport} remains valid in either case.

Equation~\eqref{eq:transport} may therefore be viewed as a retarded
quantum generalization of the classical Einstein--Helfand
construction~\cite{Einstein1905,Helfand1960}. In the classical limit and
for a diagonal dissipative channel, the canonical correlation reduces to
an ordinary equilibrium correlation, and Eq.~\eqref{eq:transport} becomes
the usual long-time growth formula for the variance of a Helfand moment,
namely the time integral of the corresponding flux. For cross and
transverse responses, the relevant long-time coefficient is obtained from
a cross-correlation of distinct moments rather than from a variance.

The quantum formulation differs from the classical relation in three
respects: the ordinary correlation is replaced by the Kubo canonical
correlation appropriate to noncommuting moments; the endpoint generated
by integration by parts is the equal-time commutator in
Eq.~\eqref{eq:integration-by-part}, which carries the
energy-magnetization content discussed below; and the response is
extracted with the retarded prescription and the order of limits specified
above.

In the charge-only sector, the use of polarization and its fluctuations under
periodic boundary conditions is established in the modern theories of
polarization and
localization~\cite{KingSmithVanderbilt1993,Resta1998,Souza2000,Resta2005}.
Eq.~\eqref{eq:transport} applies to transport problems involving arbitrary scalar sources, including those coupled to the grand-energy density and specified spin densities. It extends the moment-residue structure to transverse Hall response and provides a contact-free treatment of heat magnetization without recourse to an additional pseudogravitational vector potential.

\subsection{Heat magnetization and current gauge}
\label{sec:heat-magnetization}

For thermal transport, the contact-free form of
Eq.~\eqref{eq:transport} requires neither an auxiliary pseudogravitational
magnetic field nor a separate magnetization
subtraction~\cite{Cooper1997}. We retain only Luttinger's scalar
gravitational potential and the grand-energy polarization generated by
that scalar source. Its time derivative gives the integrated heat current,
while the equal-time endpoint generated by the moment response cancels the
explicit contact correction before the dc limit is taken. The subtraction
of internal circulating heat currents is therefore implemented directly
within the scalar-source Kubo problem, without treating heat magnetization
as an independent response to a pseudogravitational magnetic field.

We emphasize that the contact-free formula does not discard heat magnetization. Rather, it incorporates the corresponding endpoint contribution before the dc limit is taken. To connect the present formulation with the conventional magnetization-corrected description, we choose a current representative in which the antisymmetric first moment of the heat current is identified with the grand-energy moment commutator $\hat M^K_{\alpha\beta}\equiv\hat M^{KK}_{\alpha\beta}$,
\begin{equation}
\hat M^K_{\alpha\beta}
=
\hat J^K_{\alpha;\beta}
-
\hat J^K_{\beta;\alpha}.
\label{eq:qin-current-gauge}
\end{equation}
Here $\hat J^K_{\alpha;\beta}
\equiv
\left.i\partial_{q_\beta}\hat J^K_{\alpha,\boldsymbol q}
\right|_{\boldsymbol q=0}$ is defined by the same finite-$\boldsymbol q$
moment prescription as Eq.~\eqref{eq:moment}, where
$\hat J^K_{\alpha,\boldsymbol q}$ is the Fourier component of the chosen
local heat-current representative.
 The identification in Eq.~\eqref{eq:qin-current-gauge} is a choice of local-current gauge. In this gauge, the corresponding heat-magnetization density~\cite{Cooper1997,Qin2011} is
\begin{equation}
M^Q_{\alpha\beta}(T,\mu_0)
=
\frac{1}{2V}
\big\langle
\hat M^{KK}_{\alpha\beta}
\big\rangle.
\label{eq:MQ-from-MK}
\end{equation}
The factor $1/(2V)$ converts the integrated antisymmetric tensor moment into a magnetization density. This identification is a statement about the chosen current representative, not an additional transport assumption. A change of local-current gauge redistributes the current-response and magnetization contributions while leaving the contact-free transport coefficient invariant.

We now show that the identification in
Eq.~\eqref{eq:MQ-from-MK} reproduces the conventional temperature
equation for heat magnetization,
\begin{equation}
R^Q_{\alpha\beta}
=
2M^Q_{\alpha\beta}
-
T\left(
\frac{\partial M^Q_{\alpha\beta}}{\partial T}
\right)_{\mu_0}.
\label{eq:RQ-temperature}
\end{equation}
In the gauge of Eq.~\eqref{eq:qin-current-gauge}, the source
$R^Q_{\alpha\beta}$ is written as the long-wavelength antisymmetric
canonical response~\cite{Qin2011}
\begin{equation}
R^Q_{\alpha\beta}
=
-\frac{\beta_0}{2iV}
\left[
\partial_{q_\alpha}
\big\langle
\hat K_{-\boldsymbol q};
\hat J^K_{\beta,\boldsymbol q}
\big\rangle_c
-
(\alpha\leftrightarrow\beta)
\right]_{\boldsymbol q\to0},
\label{eq:qin-source}
\end{equation}
with the sign fixed by the Fourier convention in
Eq.~\eqref{eq:moment}.

At fixed $\mu_0$, the operator
$\hat M^{KK}_{\alpha\beta}$ has no explicit temperature dependence.
The canonical-correlation identity therefore gives
\begin{equation}
T\left(
\frac{\partial M^Q_{\alpha\beta}}{\partial T}
\right)_{\mu_0}
=
\frac{\beta_0}{2V}
\big\langle
\hat K;
\hat M^{KK}_{\alpha\beta}
\big\rangle_c.
\label{eq:MQ-temperature-derivative}
\end{equation}
For the long-wavelength source, one uses
\begin{equation}
\left.
\partial_{q_\alpha}
\big\langle
\hat K_{-\boldsymbol q};
\hat J^K_{\beta,\boldsymbol q}
\big\rangle_c
\right|_{\boldsymbol q=0}
=
i\big\langle
\hat K_\alpha;
\hat J^K_\beta
\big\rangle_c
-
i\big\langle
\hat K;
\hat J^K_{\beta;\alpha}
\big\rangle_c.
\label{eq:long-wavelength-source-identity}
\end{equation}
After antisymmetrization, $\hat J^K_\alpha=\dot{\hat K}_\alpha$ and the
Kubo identity give
\begin{equation}
\beta_0
\left[
\big\langle
\hat K_\alpha;
\hat J^K_\beta
\big\rangle_c
-
\big\langle
\hat K_\beta;
\hat J^K_\alpha
\big\rangle_c
\right]
=
-2
\big\langle
\hat M^{KK}_{\alpha\beta}
\big\rangle.
\label{eq:moment-current-identity}
\end{equation}
The gauge condition in Eq.~\eqref{eq:qin-current-gauge} converts the
remaining antisymmetric current moment into
$\hat M^{KK}_{\alpha\beta}$. Substitution into
Eq.~\eqref{eq:qin-source} then gives
\begin{equation}
\begin{aligned}
R^Q_{\alpha\beta}
&=
\frac{1}{V}
\big\langle
\hat M^{KK}_{\alpha\beta}
\big\rangle
-
\frac{\beta_0}{2V}
\big\langle
\hat K;
\hat M^{KK}_{\alpha\beta}
\big\rangle_c
\\
&=
2M^Q_{\alpha\beta}
-
T\left(
\frac{\partial M^Q_{\alpha\beta}}{\partial T}
\right)_{\mu_0},
\end{aligned}
\end{equation}
which is Eq.~\eqref{eq:RQ-temperature}.

Thus, in the chosen local-current gauge, the equilibrium expectation of
the commutator magnetization provides a representative of the conventional
heat magnetization and reproduces its temperature equation. The
temperature equation alone leaves a possible homogeneous contribution
proportional to $T^2$; fixing this contribution requires a convention for
the absolute heat magnetization~\cite{kapustin2020}. The contact-free
transport coefficient is independent of this representational choice.

\subsection{Invariance under density improvements}

We now return to the density ambiguity identified above and show that the contact-free coefficient is invariant under admissible density improvements. A change of the local density by a total divergence,
\begin{equation}
\hat a(\boldsymbol r)
\rightarrow
\hat a'(\boldsymbol r)
=
\hat a(\boldsymbol r)
+
\partial_\gamma\hat p^a_\gamma(\boldsymbol r),
\label{eq:density-improvement}
\end{equation}
leaves the total quantity unchanged under periodic boundary conditions. The corresponding first moment transforms as
\begin{equation}
\hat A_\alpha
\rightarrow
\hat A'_\alpha
=
\hat A_\alpha
-
\hat P^a_\alpha,
\qquad
\hat P^a_\alpha
=
\hat p^a_\alpha(\boldsymbol q=0).
\end{equation}
Substitution into Eq.~\eqref{eq:transport} gives
\begin{equation}
L_{\alpha\beta}^{\prime ab}
-
L_{\alpha\beta}^{ab}
=
-L_{\alpha\beta}^{P^a b}
-
L_{\alpha\beta}^{aP^b}
+
L_{\alpha\beta}^{P^aP^b}.
\label{eq:improvement-change}
\end{equation}
The operator $\hat P^a_\alpha$ is the zero-momentum integral of a local
improvement density rather than the first moment of a conserved density.
Provided that it has no independent projection onto a conserved
polarization mode, its correlators exhibit no secular growth: in the
classification of Eq.~\eqref{eq:asymptotic}, the mixed correlators
$\langle\hat P^a_\alpha(t);\hat B_\beta\rangle_0$ and
$\langle\hat A_\alpha(t);\hat P^b_\beta\rangle_0$ are bounded
($C_0$-type), while $\langle\hat P^a_\alpha(t);\hat P^b_\beta\rangle_0$
is regular in the dc limit. All three terms on the right-hand side of
Eq.~\eqref{eq:improvement-change} are consequently annihilated by the
$\omega^2$ prefactor. The contact-free transport coefficient is
therefore invariant under density improvements satisfying this
regularity condition. Equivalently, admissible improvements
are local redistributions whose zero-momentum integrals do not themselves
overlap with a conserved polarization or another hydrodynamic mode. This
qualification is important in gapless systems, where the finite-$\boldsymbol q$
regulator should be retained until the low-energy sector has been
projected or otherwise regularized.

Equivalently, if the source is transformed consistently, the associated integrated current changes only by a total time derivative,
\begin{equation}
\hat J^a_\alpha
\rightarrow
\hat J^{a\prime}_\alpha
=
\hat J^a_\alpha
-
\dot{\hat P}^a_\alpha.
\end{equation}
The zero-frequency admittance of the additional derivative is proportional to $i\omega$ times a susceptibility that is regular in the dc limit and therefore vanishes.

On a lattice, the same freedom corresponds to redistributing a local bond or interaction energy among neighboring sites. Two such choices differ by a lattice divergence,
\begin{equation}
\hat a'_i
=
\hat a_i
+
\nabla^-_\gamma\hat p^a_{i\gamma}.
\end{equation}
With the Fourier convention
$\hat a_{\boldsymbol q}
=
\sum_i e^{-i\boldsymbol q\cdot\boldsymbol R_i}\hat a_i$,
one has
\begin{equation}
\hat a'_{\boldsymbol q}
=
\hat a_{\boldsymbol q}
+
iq_\gamma\hat p^a_{\gamma,\boldsymbol q}
+
O(q^2),
\end{equation}
and hence $\hat A'_\alpha=\hat A_\alpha-\hat P^a_\alpha$. The preceding growth-rate argument shows that the dc coefficient is independent of how local bond or interaction energies are assigned, even though the intermediate density and local-current operators depend on that assignment.

\section{Landau--Lifshitz spin waves}

We use quadratic Landau--Lifshitz spin waves to illustrate two central features of the moment formulation. First, the calculation exposes the secular-growth structure of the contact-free formula and shows how the $\omega^2$ extraction in Eq.~\eqref{eq:transport} selects the terms that survive the dc limit. Regular contributions vanish, whereas terms in which a derivative acts on a dynamical phase grow linearly in time and supply the finite residue. Second, the formulation treats different transported densities on the same
footing. Using the grand-energy moment in the response channel gives the
magnon thermal Hall coefficient, whereas replacing it by the spin-density
moment while retaining the thermal source gives the spin Nernst
coefficient.

\subsection{Bosonic BdG moments and paraunitary geometry}

Consider a magnetically ordered state with $N_s$ transverse spin-wave coordinates in each unit cell. After choosing a local frame attached to the classical spin direction, the transverse fluctuations may be represented by Holstein--Primakoff bosons $\hat b_{a,\boldsymbol k}$ and assembled into the Nambu field
\begin{equation}
    \hat\Phi_{\boldsymbol k}
    =
    (\hat b_{1,\boldsymbol k},\cdots,
    \hat b_{N_s,\boldsymbol k},
    \hat b^\dagger_{1,-\boldsymbol k},\cdots,
    \hat b^\dagger_{N_s,-\boldsymbol k})^T .
\end{equation}
The quadratic Landau--Lifshitz Hamiltonian has the bosonic BdG form
\begin{equation}
\begin{aligned}
    \hat K
    =
    \frac{1}{2}\int[\mathrm d k]\,
    \hat\Phi_{\boldsymbol k}^\dagger
    \mathsf H(\boldsymbol k)
    \hat\Phi_{\boldsymbol k},
    \\
    [\hat\Phi_i(\boldsymbol k),\hat\Phi_j^\dagger(\boldsymbol k')]
    =
    (2\pi)^d\mathsf g_{ij}\delta(\boldsymbol k-\boldsymbol k').
\end{aligned}
    \label{eq:bdg-hamiltonian}
\end{equation}
Here $\mathsf g={\rm diag}(1_{N_s},-1_{N_s})$ is the bosonic metric, and
\begin{equation}
    \int[\mathrm d k] \equiv \int_{\mathrm{BZ}} \frac{\mathrm{~d}\boldsymbol k}{(2 \pi)^d}.
\end{equation}
The normal modes obey
\begin{equation}
    \mathsf H(\boldsymbol k)|u_{s\boldsymbol k}\rangle
    =
    \lambda_{s\boldsymbol k}\mathsf g|u_{s\boldsymbol k}\rangle,
    \qquad
    \langle u_{s\boldsymbol k}|\mathsf g|u_{t\boldsymbol k}\rangle
    =
    \varsigma_s\delta_{st},
    \label{eq:bdg-eigenproblem}
\end{equation}
where $\varsigma_s={\rm sgn}\,\lambda_s$. The positive-norm modes are the physical magnon bands, with $\lambda_{n\boldsymbol k}=\epsilon_{n\boldsymbol k}>0$; the negative-norm modes are their Nambu partners. Let $\mathsf T_{\boldsymbol k}$ be the paraunitary matrix of all positive- and negative-norm eigenvectors. Then $\mathsf T_{\boldsymbol k}^\dagger\mathsf g\mathsf T_{\boldsymbol k}=\mathsf g$ and $\mathsf T_{\boldsymbol k}^{-1}=\mathsf g\mathsf T_{\boldsymbol k}^\dagger\mathsf g$. We write $\hat\Gamma_{\boldsymbol k}=\mathsf T_{\boldsymbol k}^{-1}\hat\Phi_{\boldsymbol k}$ and $\mathsf E=\mathsf T_{\boldsymbol k}^\dagger\mathsf H\mathsf T_{\boldsymbol k}$, where $\mathsf E$ contains the positive BdG energies in both Nambu blocks. The signed frequencies governing time evolution are $\lambda_s=\varsigma_s\epsilon_s$.

The bosonic BdG Berry connection is the paraunitary connection associated with the metric $\mathsf g$. With $\partial_\alpha=\partial/\partial k_\alpha$, we define
\begin{equation}
    \mathsf A_\alpha
    =
    \mathsf T_{\boldsymbol k}^{-1}i\partial_\alpha\mathsf T_{\boldsymbol k},
    \qquad
    (\mathsf A_\alpha)_{st}
    =
    i\varsigma_s
    \langle u_{s\boldsymbol k}|\mathsf g|\partial_\alpha u_{t\boldsymbol k}\rangle .
    \label{eq:bdg-berry-connection}
\end{equation}
The covariant position operator in the spin-wave band basis is $\mathsf R_\alpha=i\partial_\alpha+\mathsf A_\alpha$; only $\mathsf A_\alpha$ is a bounded Berry-connection matrix. In particle-hole block form,
\begin{equation}
    \mathsf A_\alpha
    =
    \begin{pmatrix}
        \mathsf A^{++}_\alpha & \mathsf A^{+-}_\alpha\\
        \mathsf A^{-+}_\alpha & \mathsf A^{--}_\alpha
    \end{pmatrix} .
    \label{eq:bdg-connection-blocks}
\end{equation}
The off-diagonal blocks $\mathsf A^{+-}_\alpha$ and $\mathsf A^{-+}_\alpha$ are anomalous Berry-connection matrix elements. They vanish only in a number-conserving spin-wave problem. For a generic Landau--Lifshitz problem they are genuine parts of the paraunitary band geometry and must be retained.

In the BdG energy-polarization representative,
the first grand-energy moment is represented by the differential quadratic form
\begin{equation}
\begin{aligned}
    \hat K_\alpha
    &=
    \frac{1}{2}\int[\mathrm d k]\,
    \left[
    (\mathsf R_\alpha\hat\Gamma_{\boldsymbol k})^\dagger
    \mathsf E
    \hat\Gamma_{\boldsymbol k}
    +
    \hat\Gamma_{\boldsymbol k}^\dagger
    \mathsf E
    \mathsf R_\alpha\hat\Gamma_{\boldsymbol k}
    \right],\\
    &\mathsf R_\alpha=i\partial_\alpha+\mathsf A_\alpha .
\end{aligned}
    \label{eq:spinwave-moment}
\end{equation}

\subsection{Pole extraction and Hall responses}

Equation~\eqref{eq:spinwave-moment} must be treated as a differential quadratic form rather than as an ordinary bounded matrix element. In evaluating Eq.~\eqref{eq:transport}, contributions containing only bounded bilinear insertions give regular low-frequency correlators and are annihilated by the prefactor $\omega^2$.

The pole structure is most easily organized by asking how many momentum
derivatives act on the dynamical phases
$e^{\pm i\lambda_s t/\hbar}$ of the quasiparticle operators. Each such
action produces a factor linear in time,
\begin{equation}
i\partial_\alpha e^{\pm i\lambda_s t/\hbar}
=
\mp\frac{t}{\hbar}
(\partial_\alpha\lambda_s)
e^{\pm i\lambda_s t/\hbar}.
\end{equation}
Terms in which no derivative acts on a dynamical phase remain bounded and
contribute only to $C_0$ in Eq.~\eqref{eq:asymptotic}; they are therefore
removed by the $\omega^2$ prefactor. When exactly one derivative acts on
a phase, a band-diagonal contraction produces a contribution linear in
time. These terms constitute the finite residue $C_1$ and hence determine
the Hall coefficient.

When both derivatives act on phases, the correlator grows as $t^2$ and reduces to a band-diagonal velocity--velocity correlation. For the thermal Hall channel, the absence of this $t^2$ term from the antisymmetric response follows from the symmetry of
$C^{KK}_{2,\alpha\beta}$ established in Sec.~II. For the spin--thermal
channel, the corresponding term is proportional to
$v_{s\alpha}v_{s\beta}$ multiplied by scalar spin and energy weights, and
is therefore likewise symmetric under
$\alpha\leftrightarrow\beta$. The $t^2$ sector thus belongs to the
symmetric ballistic response and drops out of the antisymmetric Hall
coefficient in both channels.

It remains to express the $t$-linear residue in geometric form. When the
momentum derivative in one moment acts on a dynamical phase, it supplies
the diagonal band velocity
$v_{s\alpha}=\hbar^{-1}\partial_\alpha\lambda_s$. The accompanying factor
from the other moment is off diagonal in the band indices and can be
written as a matrix element of $\partial_\beta\mathsf H$. For $s\neq t$
and $\gamma=\alpha,\beta$, differentiating
Eq.~\eqref{eq:bdg-eigenproblem} gives
\begin{equation}
    (\lambda_t-\lambda_s)(\mathsf A_\gamma)_{st}
    =
    i\varsigma_s
    \langle u_{s\boldsymbol k}|
    \partial_\gamma\mathsf H
    |u_{t\boldsymbol k}\rangle.
\end{equation}
This identity converts the off-diagonal matrix elements of
$\partial_\gamma\mathsf H$ into interband Berry connections; the
diagonal band velocity generated by the dynamical phase remains as a
separate factor. After antisymmetrization in $\alpha$ and $\beta$, the
band-velocity factors and interband connections combine with the explicit Berry-connection terms into a signed BdG band sum. Rewriting this sum over positive-frequency magnon bands gives the thermal Hall coefficient
\begin{equation}
\begin{aligned}
    L^{KK}_{\alpha\beta}
    =
    \frac{k_B^2T^2}{\hbar }
    \int[\mathrm d k]\sum_{s\in +}
    \big[
    c_2(n_{s\boldsymbol k})
    -
    \tfrac{\pi^2}{3}
    \big]
    \mathcal F_{\alpha\beta,s}(\boldsymbol k).
\end{aligned}
    \label{eq:matsumoto-formula}
\end{equation}
This is the thermal response coefficient associated with the generalized thermal force $\phi_\beta^K=-\partial_\beta T/T$. With the conventional definition of the thermal conductivity, $j^K_\alpha=-\kappa_{\alpha\beta}\partial_\beta T$, the physical thermal Hall conductivity is therefore $\kappa_{\alpha\beta}=L^{KK}_{\alpha\beta}/T$. Hence Eq.~\eqref{eq:matsumoto-formula} gives the standard magnon thermal Hall conductivity after division by $T$.

For an isolated positive-frequency band, the band-resolved Abelian
curvature is
\begin{equation}
    \mathcal F_{\alpha\beta,s}
    =
    \partial_\alpha \mathsf A_{\beta,ss}
    -
    \partial_\beta \mathsf A_{\alpha,ss}.
\end{equation}
We use the notation $n_{s\boldsymbol k}=n_B(\epsilon_{s\boldsymbol k})$ and define Bose thermal weights as
\begin{equation}
c_n(x)
=
\int_0^x\! \mathrm d t
\left(
\log\tfrac{1+t}{t}
\right)^n .
\end{equation}
Equation~\eqref{eq:matsumoto-formula} is the positive-band form of the signed BdG result. With our curvature convention, the bracketed thermal weight $c_2(n_{s\boldsymbol k})-\pi^2/3$ matches the Matsumoto et al. result,\cite{MatsumotoMurakami2014} up to the conventional overall sign of the Berry curvature and the relation between $L^{KK}$ and $\kappa$. The constant term should be retained unless the total positive-band curvature integral vanishes.

For this illustration, we restrict to a longitudinal spin component
conserved by the Heisenberg Hamiltonian, so that
$\hat\Sigma^{S^z}_\alpha=0$ and
$\hat J^{S^z}_\alpha=\dot{\hat S}^z_\alpha$. Nonconserved spin channels
require the source-response terms in
Eq.~\eqref{eq:general-retarded-response}, whose explicit evaluation is
beyond the scope of the present example.
For the spin-thermal response, we specialize to the longitudinal component of Landau--Lifshitz spin waves of a $U(1)$-symmetric local-moment Heisenberg ferromagnet. We write $q_S$ for the positive spin quantum by which a positive-frequency magnon reduces the chosen spin component. Thus, for a ferromagnet polarized along $+z$, $\hat S^z(\boldsymbol r)=S_0-q_S\sum_a\hat b_a^\dagger(\boldsymbol r)\hat b_a(\boldsymbol r)$ with $q_S=\hbar$ in angular-momentum units. In the formula below we use this angular-momentum convention. Applying the finite-$\boldsymbol q$ moment definition to this density gives $\hat S^{z}_\alpha$. Inserting it into Eq.~\eqref{eq:transport} gives the spin Nernst coefficient
\begin{equation}
    L^{S^zK}_{\alpha\beta}
    =
   k_B T
    \int[\mathrm d k]\sum_{s\in+}
    c_1(n_{s\boldsymbol k})
    \mathcal F_{\alpha\beta,s}.
    \label{eq:spin-nernst}
\end{equation}
Here $L^{S^zK}_{\alpha\beta}$ is defined with respect to the generalized thermal force $\phi_\beta^K=-\partial_\beta T/T$. With the conventional definition $j^{S^z}_\alpha=-\alpha^s_{\alpha\beta}\partial_\beta T$, the physical spin Nernst coefficient is therefore
\begin{equation}
    \alpha^s_{\alpha\beta}
    =
    \frac{L^{S^zK}_{\alpha\beta}}{T}.
\end{equation}
For the present ferromagnetic benchmark, this has the Berry-curvature form characteristic of magnon spin Nernst responses discussed in Refs.~\cite{Cheng2016,Zyuzin2016,Kovalev2016}. The thermal Hall and spin Nernst responses are different components of the same moment-response framework. They share the thermal source and paraunitary band geometry, but differ in the transported moment and the corresponding Bose thermal weight.

Gapless Goldstone modes require the finite-$\boldsymbol q$ regulator in the definition of the polarization moments to be kept until the end. The exact zero mode is a collective rotation of the ordered state and should be removed in finite volume or regulated by an infinitesimal symmetry-breaking field before the thermodynamic limit is taken. For an acoustic branch, the heat-polarization matrix element carries an energy prefactor and is therefore less singular than the position matrix itself. In the Berry-curvature representation the remaining infrared behavior is controlled by the curvature of the acoustic subspace. The Hall response is regular when this curvature is integrable near the Goldstone point; if several Goldstone branches are degenerate, the Abelian curvature should be replaced by the non-Abelian curvature projected to the degenerate low-energy subspace. Longitudinal ballistic singularities of clean Goldstone modes are separate from this antisymmetric Hall response.

\section{Summary and discussion}

We have presented a unified scalar-source moment formulation of
linear-response transport. The basic object is the polarization moment
conjugate to the applied scalar force. The contact-free form does not
discard contact terms; it incorporates the equal-time endpoint
contribution into the moment response before the dc limit is taken. For a
specified spin-density component, the general moment construction gives
$\hat J^{S^\nu}_\alpha
=\dot{\hat S}^\nu_\alpha-\hat\Sigma^{S^\nu}_\alpha$.
The spin-wave illustration specializes to a conserved longitudinal spin
channel for which $\hat\Sigma^{S^z}_\alpha=0$.

The formulation separates three ambiguities that are often intertwined.
The first is the choice of local density representative. Densities that
differ by a total divergence describe the same integrated quantity, and
the contact-free coefficient is invariant under admissible improvements
of this kind. The second is the decomposition between current and source
for a nonconserved density: the continuity equation fixes only the
combination
$\partial_\alpha\hat j^a_\alpha-\hat\sigma^a$, so both operators must be
specified microscopically. The third is the local-current gauge. Once the
density and source have been fixed, a transverse circulating current may
be redistributed between the explicit current response and the
magnetization contribution without changing the integrated transport
coefficient.

In the local-current gauge of
Eq.~\eqref{eq:qin-current-gauge}, the antisymmetric first moment of the
heat current is represented by the commutator magnetization operator. Its
equilibrium expectation, divided by $2V$, gives a representative of the
heat-magnetization density that satisfies the conventional temperature
equation. The contact-free formula reproduces the corresponding
magnetization-corrected transport coefficient without introducing a
pseudogravitational magnetic field or adding a separate magnetization
term.

The quadratic Landau--Lifshitz spin-wave calculation illustrates how the
contact-free formula is used in practice and why its evaluation is
particularly simple. The $\omega^2$ extraction removes regular
contributions and isolates the terms that grow linearly in time when a
momentum derivative acts on a dynamical phase. The grand-energy and
longitudinal spin-density moments then yield the magnon thermal Hall and
spin Nernst responses, respectively, within the same paraunitary band
geometry. Recovery of the established formulas in these two channels
also provides a benchmark of the general formulation.

The operator formulation itself does not rely on a quadratic or
noninteracting approximation. Its application to interacting systems
requires the long-time or low-frequency behavior of polarization-moment
correlators rather than a separate construction of local magnetization
currents. Developing stable estimators for these correlators within
numerical many-body methods is therefore a natural next step. 

Nonconserved transport channels provide a second natural extension. In
the present work they are retained formally through the source-response
terms in Eq.~\eqref{eq:general-retarded-response}, while the spin-wave
illustration uses a source-free longitudinal spin component. Evaluating
these terms explicitly in systems with spin--orbit coupling would test
how the source--moment organization treats dissipative spin transport
beyond the conserved limit.

\acknowledgments{
The authors thank J. Shi for useful discussion. This work was supported by the National Natural Science Foundation of China (Grants No. 12274003 and No. 11725415), the National Key R\&D Program of China (Grant No. 2021YFA1400100), and the Innovation Program for Quantum Science and Technology (Grant No. 2021ZD0302600).
}

\bibliography{contact_free}

\end{document}